# Multispectral radiation temperature inversion based on Transformer-LSTM-SVM

Ying Cui[1], Kongxin Qiu[1], Shan Gao[1*], Hailong Liu[1], Rongyan Gao[1], Liwei Chen[1], Zezhan Zhang[2*],

Jing Jiang[2], Yi Niu[2], Chao Wang[3]

[1] Key Laboratory of Advanced Marine Communication and Information Technology, Ministry of Industry and Information Technology, School of Information and Communication Engineering, Harbin Engineering University, 150001 Harbin, China
[2] Clean Energy Materials and Engineering Center, School of Electronic Science and Engineering, State Key Laboratory of Electronic Thin Film and Integrated Devices, University of Electronic Science and Technology of China, 611731 Chengdu, China
[3] State Key Laboratory of Precision Measurement Technology and Instruments, Department of Precision Instrument, Tsinghua University, 100084, Beijing, China

**Abstract:** The key challenge in multispectral radiation thermometry is accurately measuring emissivity. Traditional constrained optimization methods often fail to meet practical requirements in terms of precision, efficiency, and noise resistance. However, the continuous advancement of neural networks in data processing offers a potential solution to this issue. This paper presents a multispectral radiation thermometry algorithm that combines Transformer, LSTM (Long Short-Term Memory), and SVM (Support Vector Machine) to mitigate the impact of emissivity, thereby enhancing accuracy and noise resistance. In simulations, compared to the BP neural network algorithm, GIM-LSTM, and Transformer-LSTM algorithms, the Transformer-LSTM-SVM algorithm demonstrates an improvement in accuracy of 1.23%, 0.46% and 0.13%, respectively, without noise. When 5% random noise is added, the accuracy increases by 1.39%, 0.51%, and 0.38%, respectively. Finally, experiments confirmed that the maximum temperature error using this method is less than 1%, indicating that the algorithm offers high accuracy, fast processing speed, and robust noise resistance. These characteristics make it well-suited for real-time high-temperature measurements with multi-wavelength thermometry equipment.

---

[*] Corresponding author.
E-mail addresses: gaoshan08@hrbeu.edu.cn (Shan Gao)
E-mail addresses: zhangzz@uestc.edu.cn (Zezhan Zhang)

# 1. Introduction

Multispectral thermometry is a vital radiation temperature measurement method. It measures the radiation energy of an object at multiple wavelengths and uses mathematical models for inversion calculations to obtain the object's temperature distribution[1-4]. However, due to the influence of unknown emissivity, the inversion process often requires prior assumptions about the emissivity model based on experience. Moreover, obtaining the emissivity function is challenging, which affects the accuracy of temperature measurement results[5, 6]. Therefore, how to estimate temperature without relying on assumed emissivity models remains a key focus of current research.

To overcome the dependency on the assumed model between the material emissivity and wavelength, in 1998, Xiaogang Sun first applied neural network theory to the field of multispectral radiation thermometry, effectively overcoming the dependence on the emissivity model and providing a new method for measuring actual temperature of targets[7]. In 2020, Gaoshan used a BP network to adaptively establish the emissivity model, construct constraint equations, and solve the proper temperature using an improved non-dominated sorting genetic algorithm[8]. However, due to the drawbacks of BP neural networks, such as being prone to local minima, slow convergence, and weak generalization ability, the above two methods struggle to meet practical application requirements[9]. In 2023, Xing Jian proposed a Long Short-Term Memory(LSTM) multispectral radiation measurement algorithm that considers the temporal continuity of multispectral radiation data[10]. However, since the data is only learned through the LSTM network, significant limitations, such as weak feature extraction capabilities, affect the prediction accuracy. These studies suggest that neural network technology holds great promise for multispectral radiation temperature measurement, particularly in mitigating the effects of the emissivity model. However, these methods still face challenges in practical applications, such as accuracy in prediction and sensitivity to noise, which require further research and refinement.

This paper utilizes a combined Transformer and LSTM model for feature processing. The Transformer effectively captures complex feature relationships, while LSTM handles temporal dependencies. Additionally, SVM's powerful nonlinear processing capabilities and noise robustness are leveraged for regression prediction. As a result, this paper integrates the Transformer-LSTM neural network algorithm with SVM for accurate temperature inversion in radiation thermometry, enhancing the measurement process's precision and efficiency.

# 2. Basic principles

For multi-wavelength temperature measurement devices, the output signal value of the $i$th channel is $V_i$, as shown in Equation (1):

$$V_i = A_{\lambda_i} \cdot \varepsilon(\lambda_i, T) \cdot \lambda_i^{-5} \cdot e^{-\frac{C_2}{\lambda_i T}} \quad (i=1,2,...,n) \quad (1)$$

In the equation, $A_{\lambda_i}$ is the verification constant that is only related to the wavelength and independent of temperature $\varepsilon(\lambda_i, T)$ is the spectral emissivity of the target at temperature $T$, and $C_2$ is the second radiation constant[11, 12].

When the reference temperature of the determined blackbody is set to $T'$, the emissivity of the blackbody at this moment is 1, the output signal of its $i$th channel is $V_i'$, as shown in Equation (2).

$$V_i' = A_{\lambda_i} \cdot \lambda_i^{-5} \cdot e^{-\frac{C_2}{\lambda_i T'}} \quad (2)$$

From Equations (1) and (2), Equation (3) can be derived:

$$\frac{V_i}{V_i'} = \varepsilon(\lambda_i, T) \cdot e^{-\frac{C_2}{\lambda_i T}} \cdot e^{\frac{C_2}{\lambda_i T'}} \quad (3)$$

This ratio $\dfrac{V_i}{V_i'}$ reflects the target's radiative energy change at temperatures different from the reference temperature.

The key issue in multispectral radiation thermometry is how to measure emissivity accurately. Suppose the radiation temperature measurement method assumes an emissivity model in advance. In that case, there may be a mismatch between the actual emissivity and the predicted model, leading to significant temperature measurement errors[13, 14]. To address the issue of the indeterminacy of the emissivity model, a hybrid model called Transformer-LSTM-SVM is proposed. This hybrid model efficiently handles complex data by combining the strengths of different models while eliminating the dependence on the emissivity model. The data collected by the multispectral radiation thermometry equipment comes from the surface radiation of the object. The radiation intensity rise process can be seen as an ordered sequence. Transformer-LSTM is especially good at handling sequential data and can extract richer features from the input data, and the feature extraction process is shown in Fig 1.

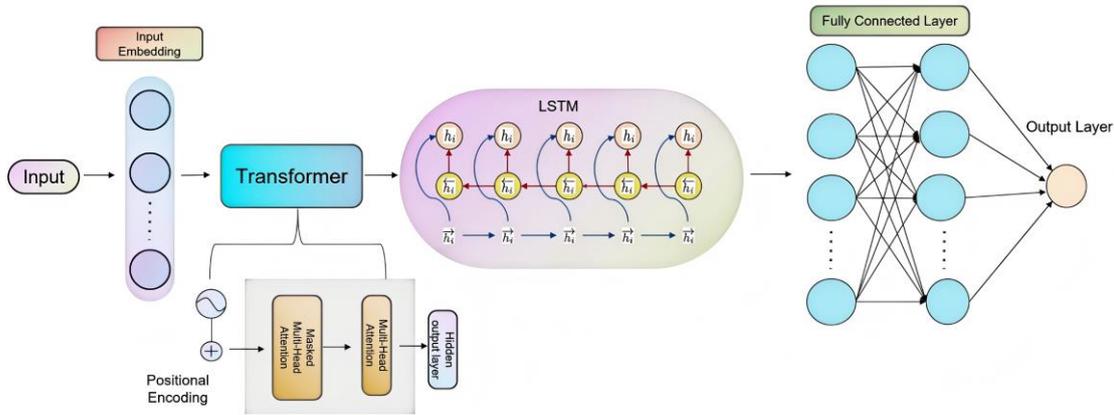

Fig.1. Transformer-LSTM Model.

The features extracted by the Transformer-LSTM model are used as input to the SVM model, which better fits the relationship between the radiative energy ratio and temperature. Fig 2 shows the flowchart of the multispectral infrared radiation thermometry algorithm based on the Transformer-LSTM-SVM model.

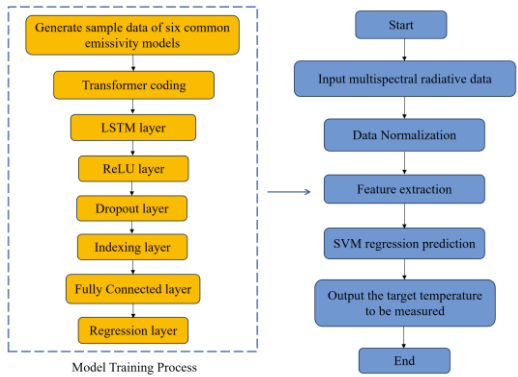

Fig.2. Flowchart of the multispectral infrared radiation thermometry algorithm based on the Transformer-LSTM-SVM model.

## 3. Simulation experiments

3.1 Preparation of dataset

A dataset uses the above ratios as inputs and the target temperature as the output. This dataset contains the spectral radiative energy ratios at multiple temperature points and their corresponding true temperatures. This paper uses a six-wavelength radiative temperature measurement as an example. According to the principle of multispectral radiation thermometry, in high-temperature regions, the energy of blackbody radiation is primarily concentrated in the short-wavelength region[15]. Therefore, the experimental wavelengths selected are 1.5μm, 1.6μm, 1.7μm, 1.8μm, 1.9μm, and 2.0μm. The data collected at these six wavelengths can better reflect the temperature changes on the object's surface.

First, we set the temperature range from 923K to 1253K, with a temperature point every 10K, totaling 34 temperature points, with a reference temperature of 848K. The simulation uses six common emissivity types: A, B, C, D, E, and F, as shown in Fig 3. A combination of 60 emissivities is designed under each emissivity model to make the experimental data more general, resulting in 34*60*6 data sets. 11,016 data points are randomly selected for network training, and 1,224 are reserved for testing.

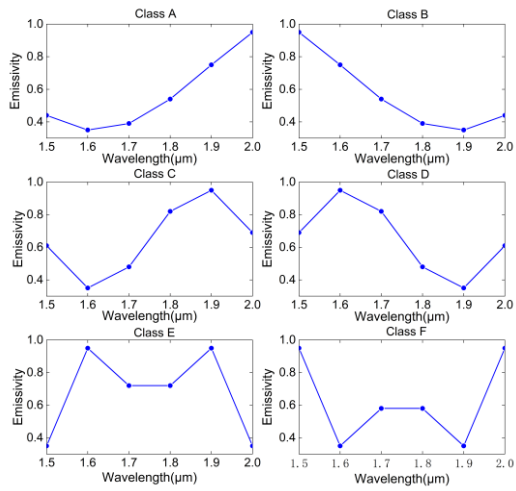

Fig. 3. Six common emissivity models.

### 3.2 Simulation results

To verify the accuracy and effectiveness of the method, the Transformer-LSTM-SVM network is compared with different types of neural networks. The BP neural, GIM-LSTM, Transformer-LSTM, and Transformer-LSTM-SVM neural networks are trained and tested separately. The simulation results are shown in Fig 4. The average relative temperature prediction errors for the six emissivity models using the Transformer-LSTM-SVM algorithm, BP algorithm, GIM-LSTM algorithm, and Transformer-LSTM algorithm are 0.11%, 1.34%, 0.57%, and 0.24%, respectively. Therefore, the performance of the Transformer-LSTM-SVM algorithm is superior to other traditional methods.

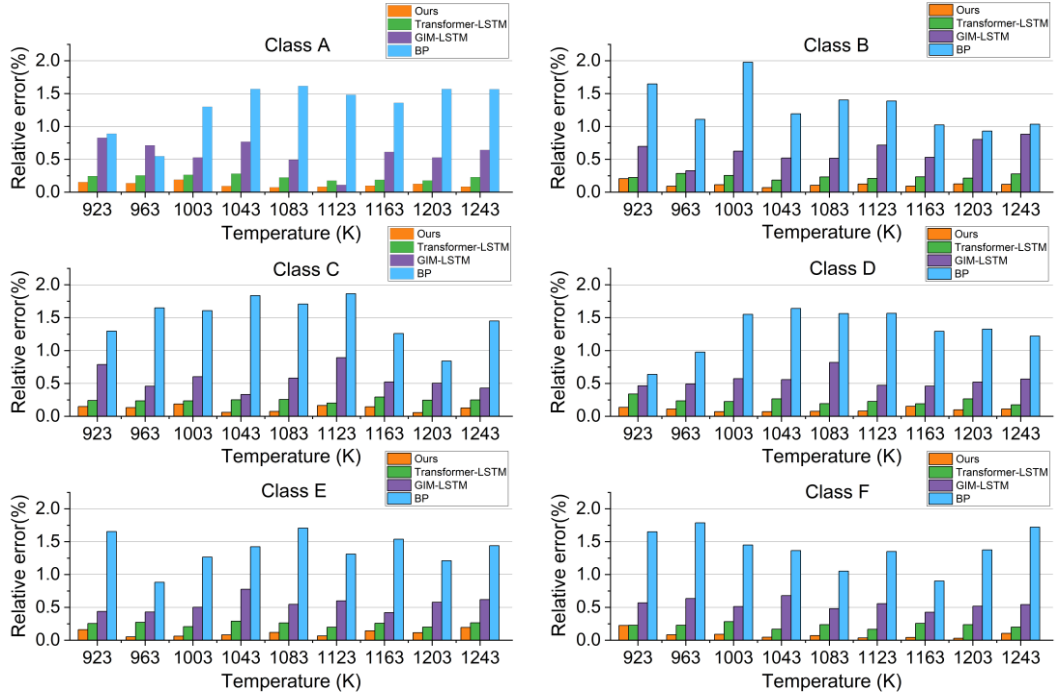

Fig. 4. Comparison of temperature inversion errors for different network model

## 3.3 Simulation of anti-noise performance

In practical applications, the multi-wavelength radiation thermometry process can be affected by noise, leading to data distortion. Therefore, to study the impact of random noise on the method, Gaussian random noise at a 5% level is added to the input data of the dataset. The samples with Gaussian random noise are then input into the neural network models designed in the above experiment. The results are shown in Fig 5. From the results in Fig 5, it can be seen that the average relative temperature prediction error for the six emissivity models using the Transformer-LSTM-SVM algorithm is 1.04%, for the BP algorithm is 2.43%, GIM-LSTM algorithm is 1.55%, and for the Transformer-LSTM algorithm is 1.42%. Therefore, compared to the literature's BP neural network[7] and GIM-LSTM methods[10], the Transformer-LSTM-SVM method exhibits better adaptability to noisy environments. At the same time, a comparison between the Transformer-LSTM-SVM model and the Transformer-LSTM model shows that the model with the added SVM module has better noise resistance performance.

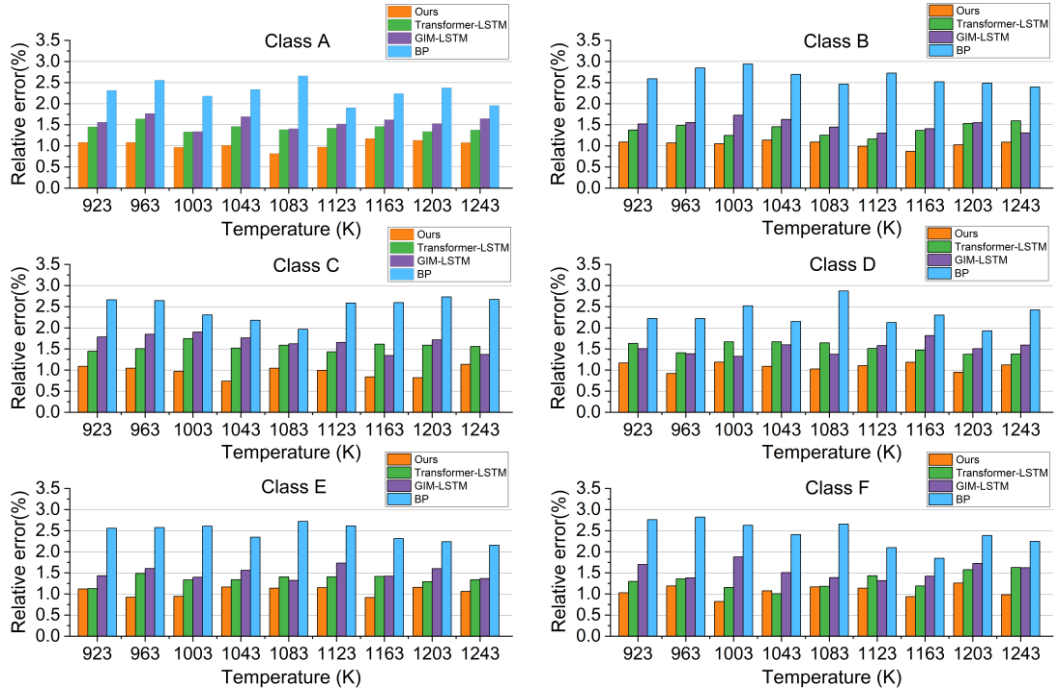

Fig. 5. Comparison of errors for different network models with added 5% noise.

3.4 Validation of experimental data

To verify the effectiveness of the algorithm proposed in this paper, an experimental setup for measuring the temperature of GH3044 alloy material was designed, as shown in Fig 6. The system's front end consists of a collimating lens and a spectrometer. The collimating lens is responsible for collecting the spectral radiation information of the material and transmitting the collected radiation data to the spectrometer via optical fibers[16]. The spectrometer divides the incoming spectral radiation information into different wavelength bands. The spectral radiation data for each corresponding wavelength channel is obtained through spectrometry. Finally, the radiation data is input into the terminal for processing and calculation, resulting in the corresponding real temperature.

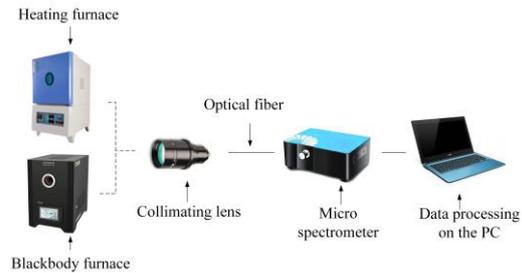

Fig. 6. Schematic diagram of the experimental setup.

The experimental temperature range is selected from 923 K to 1253 K, with a reference temperature of 848 K. The verification results are shown in Fig 7. The figure shows that the error variation of the simulated target at nine temperature points is relatively smooth, and the average error of the validation experiment is less than 1%, sufficient to meet the practical measurement needs of most industrial scenarios, thus proving the algorithm's effectiveness.

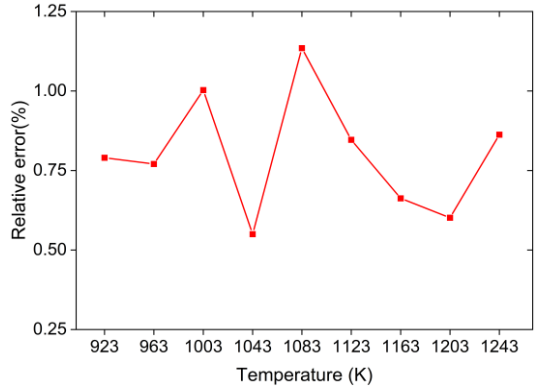

Fig. 7. Relative error of temperature inversion in the verification experiment.

In addition, the running time of the new algorithm is relatively faster than that of other algorithms. The computation time of the Transformer-LSTM-SVM algorithm is 0.035 seconds. In comparison, the average running time of the BP algorithm is 0.11 seconds, the average running time of the GIM-LSTM algorithm is 0.061 seconds，and the Transformer-LSTM algorithm is 0.041 seconds. The Transformer-LSTM-SVM algorithm uses Transformer-LSTM for training and feature extraction and SVM for prediction, leveraging SVM's strong nonlinear processing capability. It typically requires only simple kernel function calculations for feature mapping, resulting in a significantly lower time cost during the prediction phase.

## 4. Conclusions

This study combines the Transformer-LSTM neural network algorithm with the SVM algorithm to process data from multi-wavelength pyrometers. The results show that this integrated method not only ensures high accuracy in temperature measurements but also significantly improves data processing efficiency, thereby reducing the impact of unknown emissivity on the quality of temperature measurements for different materials. The simulation results show that this method has high radiation temperature measurement accuracy. Compared to the traditional BP neural network algorithm, GIM-LSTM, and Transformer-LSTM algorithms, the noise-free Transformer-LSTM-SVM algorithm can improve accuracy by 1.23%, 0.46%, and 0.13%, respectively. After adding 5% noise, the accuracy of the Transformer-LSTM-SVM algorithm is higher than that of the BP neural network algorithm by 1.39%, the GIM-LSTM algorithm by 0.51%, and the Transformer-LSTM algorithm by 0.38%. Finally, experimental verification shows that the maximum temperature error of this method is less than 1%, which can meet the temperature measurement requirements in most cases. In conclusion, this algorithm has the advantages of high precision, rapid computation and good noise resistance, making it suitable for real-time high-temperature measurement with multi-wavelength thermometry equipment.

**CRediT authorship contribution statement**

Ying Cui: Writing – review & editing, Writing – original draft, Project administration, Methodology. Kongxin Qiu: Writing – review & editing, Supervision. Shan Gao: Writing – review & editing, Supervision Funding acquisition, Data curation, Conceptualization. Hailong Liu: Validation, Conceptualization. Rongyan Gao: Validation, Conceptualization. Liwei Chen: Validation. Zezhan Zhang: Writing – review & editing, Supervision Funding acquisition, Data curation, Conceptualization. Jing Jiang: Validation. Yi Niu: Validation. Chao Wang: Funding acquisition.

**Declaration of competing interest**

The authors declare that they have no known competing financial interests or personal relationships that could have appeared to influence the work reported in this paper.

**Data availability**

Data will be made available on request.

**Acknowledgments**

This work is supported by the National Natural Science Foundation of China (62275059,62225406); Heilongjiang Provincial Natural Science Foundation of China (No. YQ2023F014) ；State Key Laboratory of Precision Measurement Technology and Instruments (2024PMTI03); Sichuan Science and Technology Program under Grant Number 24MZGC0019.